\def\complexNumbers{\mathbb{C}}
\def\realNumbers{\mathbb{R}}

\def\constante{{\rm e}}

\def\expectationOperator[#1][#2]{{\mathbb{E}_{#2}}\left[#1\right]}
\def\indicatorFunction[#1]{\mathbb{I}\left[{#1}\right]}
\def\probability[#1]{\mathbb{P}\left[{#1}\right]}

\def\numberOfEdgeDevices{K}
\def\timeDomainOFDM[#1]{x(#1)}
\def\timeVar{t}

\def\numberOfActiveSubcarriers{M}
\def\idftSize{N_{\rm IDFT}}
\def\dataSymbols[#1]{d_{#1}}

\def\symbolDuration{T_{\rm s}}
\def\Ptransmit{P_{\rm tx}}
\def\receivedSymbolAtSubcarrier[#1]{r_{#1}}
\def\transmittedSymbolAtSubcarrier[#1]{t_{#1}}
\def\randomSymbolAtSubcarrier[#1]{s_{#1}^{(\indexCommunicationRound)}}
\def\channelAtSubcarrier[#1]{h_{#1}}
\def\noiseAtSubcarrier[#1]{n_{#1}}
\def\numberOfOFDMSymbols{S}
\def\indexOFDMSymbol{m}
\def\asymbolFromED[#1]{d_{#1}}

\def\exponentialIntegral[#1]{{\rm Ei}(#1)}
\def\tciFactor[#1]{p_{#1}}
\def\mappingFunction{{f}}
\def\encoder[#1]{\psi_#1}
\def\symbolEnergy{E_{\rm s}}

\def\voteInTime[#1]{m^{#1}}
\def\voteInFrequency[#1]{l^{#1}}

\def\numberOFEDsForOptionOne{K^{+}_{\indexGradient}}
\def\numberOFEDsForOptionSecond{K^{-}_{\indexGradient}}
\def\noiseVariance{\sigma_{\rm n}^2}

\def\coefficientOne{a}

\def\correctDecision[#1]{p_{#1}}
\def\incorrectDecision[#1]{q_{#1}}
\def\aparameterForBer[#1]{\epsilon_{#1}}
\def\numberOfEDsWithCorrectChoice{Z}
\def\probabilityIncorrect[#1]{P^{\rm err}_{#1}}
\def\meanOptionOne{\mu^{+}_{\indexGradient}}
\def\meanOptionTwo{\mu^{-}_{\indexGradient}}
\def\effectiveSNR{\xi}

\def\identityMatrix[#1]{\textbf{\textrm{I}}_{#1}}
\def\zeroVector[#1]{\textbf{\textrm{0}}_{#1}}

\def\dataset[#1]{\mathcal{D}_{#1}}
\def\cardinalityOfLocalData{D}
\def\datasetBatch[#1]{\mathcal{\tilde{D}}_{#1}}
\def\batchSize{n_{\rm b}}
\def\brachSizeRelativeToRounds{\gamma}
\def\completeData{\mathcal{D}}
\def\numberOfModelParameters{q}
\def\sampleData[#1]{{\textrm{\textbf{x}}}_{#1}}
\def\sampleLabel[#1]{{y}_{#1}}
\def\learningRate{\eta}

\def\deltaVectorAtIteration[#1][#2]{{\Delta}^{(#1)}_{#2}}
\def\deltaVectorAtIterationEle[#1]{{\bm \Delta}^{#1}}

\def\indexED{k}
\def\indexGradient{i}
\def\indexSampleData{{\ell}}
\def\indexCommunicationRound{n}

\def\modelParametersAtIteration[#1]{\textbf{w}^{(#1)}}
\def\modelParametersAtIterationEle[#1][#2]{w^{(#1)}_{#2}}

\def\modelParameters{\textbf{w}}
\def\modelParametersEle[#1]{{w}_{#1}}
\def\modelParametersOptimal{{\textbf{w}^{*}}}

\def\localGradientSign[#1][#2]{\bar{\textbf{g}}_{#1}^{(#2)}}
\def\localGradient[#1][#2]{\tilde{\textbf{g}}_{#1}^{(#2)}}
\def\localGradientNoIndex[#1]{\tilde{\textbf{g}}_{#1}}

\def\localGradientSignElement[#1][#2]{{\bar{g}}_{#1}^{(#2)}}
\def\localGradientElement[#1][#2]{{\tilde{g}}_{#1}^{(#2)}}
\def\localGradientNoIndexElement[#1]{{\tilde{g}}_{#1}}

\def\lossFunctionSample[#1]{f(#1)}
\def\lossFunctionLocal[#1][#2]{F_{#1}(#2)}
\def\lossFunctionGlobal[#1]{F(#1)}
\def\lossFunctionGlobalMinimum{F^*}

\def\majorityVoteEle[#1][#2]{{v}^{(#1)}_{#2}}
\def\majorityVote[#1]{\textbf{v}^{(#1)}}

\def\globalGradient[#1]{{\textbf{{g}}}^{(#1)}}
\def\globalGradientElement[#1][#2]{{{g}}^{(#1)}_{#2}}
\def\globalGradientNoIndex{{\textbf{{g}}}}
\def\globalGradientElementNoIndex[#1]{{g_{#1}}}

\def\communicationRounds{N}

\def\metricForFirst[#1]{e_{#1}^{+}}
\def\metricForSecond[#1]{e_{#1}^{-}}

\def\nonnegativeConstants{\textbf{L}}
\def\nonnegativeConstantsEle[#1]{L_{#1}}
\def\varianceBound{{\bm \sigma}}
\def\varianceBoundEle[#1]{\sigma_{#1}}

\def\rmsDelaySpread{T_{\rm rms}}

\def\symbolVector[#1]{\textbf{\textrm{d}}_{#1}^{(\indexCommunicationRound)}}
\def\symbolVectorEstimate[#1]{\tilde{\textbf{\textrm{{d}}}}_{#1}^{(\indexCommunicationRound)}}
\def\receivedVector[#1]{\textbf{\textrm{r}}_{#1}^{(\indexCommunicationRound)}}
\def\noiseVector[#1]{\textbf{\textrm{n}}_{#1}^{(\indexCommunicationRound)}}

\def\transmittedVector[#1]{\textbf{\textrm{t}}_{#1}^{(\indexCommunicationRound)}}
\def\idftMatrix[#1]{\textbf{\textrm{F}}_{#1}^{\rm H}}
\def\dftMatrix[#1]{\textbf{\textrm{F}}_{#1}}
\def\transformPrecoder[#1]{\textbf{\textrm{D}}_{#1}}
\def\transformDecoder[#1]{\textbf{\textrm{D}}_{#1}^{\rm H}}
\def\frequencyMapping{\textbf{\textrm{M}}_{\textrm{f}}}

\def\channelMatrix[#1]{\textbf{\textrm{H}}_{#1}^{(\indexCommunicationRound)}}

\def\samplePeriod{T_{\rm sample}}
\def\symbolSpacing{T_{\rm spacing}}

\def\pulseDuration{T_{\rm pulse}}
\def\syncError{T_{\rm sync}}
\def\channelSpread{T_{\rm chn}}

\def\guardTimeUniform{T_{\rm gap}}
\def\numberOfVotesPerDFTsOFDM{M_{\rm vote}}
\def\uniformGap{M_{\rm gap}}
\def\numberOfElementsInAPulse{M_{\rm pulse}}
\def\symbolsActivatedBins{\textbf{p}}

\newcommand\mydots{\hbox to 1em{.\hss.\hss.}}
\documentclass[conference,comsoc]{IEEEtran}
\usepackage{multicol}
\usepackage{lipsum}
\usepackage{mathtools}
\usepackage{amsthm}
\usepackage{acronym}
\usepackage{stfloats}
\usepackage{amsfonts}
\usepackage{acronym}
\usepackage{cite}
\usepackage{multirow}
\usepackage{bm}
\usepackage{amsmath,amssymb}
\usepackage{graphicx}
\usepackage{epstopdf}
\usepackage[table,xcdraw]{xcolor}
\usepackage[caption=false,font=footnotesize]{subfig}
\usepackage[geometry]{ifsym}
\usepackage{array}
\usepackage[utf8]{inputenc}
\usepackage[T1]{fontenc}
\let\norm\undefined % <-- "Undefine" \norm
\DeclarePairedDelimiter\norm{\lVert}{\rVert}

\usepackage{tikz}
\usepackage{lipsum}
\usetikzlibrary{calc,shadings,patterns}
\makeatletter
\tikzset{%
  remember picture with id/.style={%
    remember picture,
    overlay,
    save picture id=#1,
  },
  save picture id/.code={%
    \edef\pgf@temp{#1}%
    \immediate\write\pgfutil@auxout{%
      \noexpand\savepointas{\pgf@temp}{\pgfpictureid}}%
  },
  if picture id/.code args={#1#2#3}{%
    \@ifundefined{save@pt@#1}{%
      \pgfkeysalso{#3}%
    }{
      \pgfkeysalso{#2}%
    }
  }
}

\def\savepointas#1#2{%
  \expandafter\gdef\csname save@pt@#1\endcsname{#2}%
}

\def\tmk@labeldef#1,#2\@nil{%
  \def\tmk@label{#1}%
  \def\tmk@def{#2}%
}

\tikzdeclarecoordinatesystem{pic}{%
  \pgfutil@in@,{#1}%
  \ifpgfutil@in@%
    \tmk@labeldef#1\@nil
  \else
    \tmk@labeldef#1,(0pt,0pt)\@nil
  \fi
  \@ifundefined{save@pt@\tmk@label}{%
    \tikz@scan@one@point\pgfutil@firstofone\tmk@def
  }{%
  \pgfsys@getposition{\csname save@pt@\tmk@label\endcsname}\save@orig@pic%
  \pgfsys@getposition{\pgfpictureid}\save@this@pic%
  \pgf@process{\pgfpointorigin\save@this@pic}%
  \pgf@xa=\pgf@x
  \pgf@ya=\pgf@y
  \pgf@process{\pgfpointorigin\save@orig@pic}%
  \advance\pgf@x by -\pgf@xa
  \advance\pgf@y by -\pgf@ya
  }%
}

\makeatother
\newcounter{hatchNumber}
\DeclarePairedDelimiter\ceil{\lceil}{\rceil}
\DeclarePairedDelimiter\floor{\lfloor}{\rfloor}

\usepackage{cuted}
\makeatletter
\newif\ifAC@uppercase@first%
\def\Aclp#1{\AC@uppercase@firsttrue\aclp{#1}\AC@uppercase@firstfalse}%
\def\AC@aclp#1{%
	\ifcsname fn@#1@PL\endcsname%
	\ifAC@uppercase@first%
	\expandafter\expandafter\expandafter\MakeUppercase\csname fn@#1@PL\endcsname%
	\else%
	\csname fn@#1@PL\endcsname%
	\fi%
	\else%
	\AC@acl{#1}s%
	\fi%
}%
\def\Acp#1{\AC@uppercase@firsttrue\acp{#1}\AC@uppercase@firstfalse}%
\def\AC@acp#1{%
	\ifcsname fn@#1@PL\endcsname%
	\ifAC@uppercase@first%
	\expandafter\expandafter\expandafter\MakeUppercase\csname fn@#1@PL\endcsname%
	\else%
	\csname fn@#1@PL\endcsname%
	\fi%
	\else%
	\AC@ac{#1}s%
	\fi%
}%
\def\Acfp#1{\AC@uppercase@firsttrue\acfp{#1}\AC@uppercase@firstfalse}%
\def\AC@acfp#1{%
	\ifcsname fn@#1@PL\endcsname%
	\ifAC@uppercase@first%
	\expandafter\expandafter\expandafter\MakeUppercase\csname fn@#1@PL\endcsname%
	\else%
	\csname fn@#1@PL\endcsname%
	\fi%
	\else%
	\AC@acf{#1}s%
	\fi%
}%
\def\Acsp#1{\AC@uppercase@firsttrue\acsp{#1}\AC@uppercase@firstfalse}%
\def\AC@acsp#1{%
	\ifcsname fn@#1@PL\endcsname%
	\ifAC@uppercase@first%
	\expandafter\expandafter\expandafter\MakeUppercase\csname fn@#1@PL\endcsname%
	\else%
	\csname fn@#1@PL\endcsname%
	\fi%
	\else%
	\AC@acs{#1}s%
	\fi%
}%
\edef\AC@uppercase@write{\string\ifAC@uppercase@first\string\expandafter\string\MakeUppercase\string\fi\space}%
\def\AC@acrodef#1[#2]#3{%
	\@bsphack%
	\protected@write\@auxout{}{%
		\string\newacro{#1}[#2]{\AC@uppercase@write #3}%
	}\@esphack%
}%
\def\Acl#1{\AC@uppercase@firsttrue\acl{#1}\AC@uppercase@firstfalse}
\def\Acf#1{\AC@uppercase@firsttrue\acf{#1}\AC@uppercase@firstfalse}
\def\Ac#1{\AC@uppercase@firsttrue\ac{#1}\AC@uppercase@firstfalse}
\def\Acs#1{\AC@uppercase@firsttrue\acs{#1}\AC@uppercase@firstfalse}

\newtheorem{theorem}{Theorem}

\newtheorem{assumption}{Assumption}

\DeclareMathOperator{\sign}{sign}
\def\signNormal[#1]{\sign\left(#1\right)}
\acrodef{SNR}{signal-to-noise ratio}
\acrodef{RMSE}{root-mean-square error}
\acrodef{OFDM}{orthogonal frequency division multiplexing}
\acrodef{DFT}{discrete Fourier transform}
\acrodef{PSK}{phase-shift keying}
\acrodef{QAM}{quadrature amplitude modulation}
\acrodef{QPSK}{quadrature phase-shift keying}
\acrodef{PMEPR}{peak-to-mean envelope power ratio}
\acrodef{BER}{bit-error ratio}
\acrodef{SNR}{signal-to-noise ratio}
\acrodef{PSD}{power spectral density}
\acrodef{SE}{spectral efficiency}
\acrodef{CP}{cyclic prefix}
\acrodef{AWGN}{additive white Gaussian noise}
\acrodef{CFR}{channel frequency response}
\acrodef{CIR}{channel impulse response}
\acrodef{MMSE}{minimum mean square error}
\acrodef{LMMSE}{linear minimum mean square error}
\acrodef{BPSK}{binary phase shift keying}
\acrodef{BLER}{block-error rate}
\acrodef{ML}{maximum likelihood}
\acrodef{PHY}{physical layer}
\acrodef{PA}{power amplifier}
\acrodef{IDFT}{inverse DFT}
\acrodef{DoF}{degrees-of-freedom}
\acrodef{IoT}{Internet-of-Things}
\acrodef{FDE}{frequency-domain equalization}
\acrodef{RF}{radio-frequency}
\acrodef{IM}{index modulation}
\acrodef{BS}{base station}
\acrodef{MF}{matched filter}
\acrodef{PPM}{pulse-position modulation}

\acrodef{PPM-MV}[PPM-MV]{\ac{PPM}-based \ac{MV}}

\acrodef{BAA}{broadband analog aggregation}
\acrodef{OBDA}{one-bit broadband digital aggregation}
\acrodef{FEEL}{federated edge learning}
\acrodef{FL}{federated learning}
\acrodef{ED}{edge device}
\acrodef{ES}{edge server}
\acrodef{UL}{uplink}
\acrodef{DL}{downlink}
\acrodef{OAC}[AirComp]{over-the-air computation}
\acrodef{TCI}{truncated-channel inversion}
\acrodef{MV}{majority vote}
\acrodef{CNN}{convolutional neural network}
\acrodef{ReLU}{rectified-linear unit}
\acrodef{CSI}{channel state information}
\acrodef{PAPR}{peak-to-average power ratio}
\acrodef{SC}{single-carrier}
\acrodef{iid}[IID]{independent and identically distributed}
\acrodef{RMS}{root-mean-square}
\acrodef{4G}{Fourth Generation}
\acrodef{5G}{Fifth Generation}
\acrodef{NR}{New Radio}
\acrodef{LTE}{Long-Term Evolution}
\acrodef{SGD}{stochastic gradient descend}
\acrodef{signSGD}{sign stochastic gradient descend}

\acrodef{5G}{Fifth Generation}
\acrodef{4G}{Fourth Generation}
\acrodef{NR}{New Radio}
\acrodef{LTE}{Long Term Evolution}
\acrodef{PRACH}{physical random access channel}
\acrodef{PUCCH}{physical uplink control channel}

\acrodef{DFT-s-OFDM}{\ac{DFT}-spread \ac{OFDM}}

\def\BibTeX{{\rm B\kern-.05em{\sc i\kern-.025em b}\kern-.08em
    T\kern-.1667em\lower.7ex\hbox{E}\kern-.125emX}}
\begin{document}

\title{Over-the-Air Computation with DFT-spread OFDM for Federated Edge Learning}

%\author{
%\IEEEauthorblockN{Alphan \c{S}ahin\IEEEauthorrefmark{1}, Bryson Everette\IEEEauthorrefmark{1}, Safi Shams Muhtasimul Hoque\IEEEauthorrefmark{1}}
%\IEEEauthorblockA{\IEEEauthorrefmark{1}Electrical  Engineering Department,
%University of South Carolina, Columbia, SC, USA}
%Email: asahin@mailbox.sc.edu, everetb@email.sc.edu,  shoque@email.sc.edu}

\author{
	\IEEEauthorblockN{Alphan \c{S}ahin}
	\IEEEauthorblockA{Electrical  Engineering Department\\
	University of South Carolina\\
	Columbia, SC, USA\\
	Email: asahin@mailbox.sc.edu}
	\and	
	\IEEEauthorblockN{Bryson Everette}
	\IEEEauthorblockA{Electrical  Engineering Department\\
		University of South Carolina\\
		Columbia, SC, USA\\
		Email: everetb@email.sc.edu}
	\and
	\IEEEauthorblockN{Safi Shams Muhtasimul Hoque}
	\IEEEauthorblockA{Electrical  Engineering Department\\
		University of South Carolina\\
		Columbia, SC, USA\\
		Email: shoque@email.sc.edu	}

}

\maketitle

\begin{abstract}
In this study, we propose an \ac{OAC} scheme for \ac{FEEL} without \ac{CSI} at the \acp{ED} or the \ac{ES}. The proposed scheme relies on  non-coherent communication techniques for achieving distributed training  by \ac{MV}. In this work, the votes, i.e., the signs of the local gradients, from the \acp{ED} are represented with the \ac{PPM} symbols constructed with \ac{DFT-s-OFDM}. By taking the delay spread and time-synchronization errors into account, the \ac{MV} at the \ac{ES} is obtained  with an energy detector.  Hence, the proposed scheme does not require \ac{CSI} at the \acp{ED} and \ac{ES}. We also prove the convergence of the distributed training when the \ac{MV} is obtained with the proposed scheme under fading channel.
%The proposed method naturally reduces the \ac{PMEPR} of the signal as it inherits the properties of the \ac{SC} waveform. 
Through simulations, we show that the proposed scheme provides a high test accuracy in fading channels while resulting in lower \ac{PMEPR} symbols.
%, as well as how resource utilization affects the effectiveness of \ac{DFT-s-OFDM} in comparison to \ac{OBDA} with \ac{QAM}. 
\end{abstract}

%\begin{IEEEkeywords}
%DFT-s-OFDM, distributed learning, federated edge learning, PPM,  over-the-air computation, PMEPR.
%\end{IEEEkeywords}
\section{Introduction}

\acresetall
\Ac{FEEL} is an implementation of \ac{FL} over a wireless network to train a model by using the local data at the \acp{ED} without uploading them to an \ac{ES} \cite{gafni2021federated, chen2021distributed}. Within each iteration of \ac{FEEL}, a substantial number of parameters (e.g., model parameters or model updates) from each \ac{ED} needs to be transmitted to the \ac{ES} for aggregation. Thus, the communication aspect of \ac{FEEL} is one of the major bottlenecks. One of the promising solutions to this issue is to perform the aggregation by utilizing the signal-superposition property of a wireless multiple access channel \cite{Goldenbaum_2013,Wanchun_2020,Nazer_2007}, i.e., \ac{OAC}. However, an \ac{OAC} scheme often requires \ac{CSI} at either the \acp{ED} or \ac{ES} to maintain coherent superposition of the signals from \acp{ED}, which can cause a non-negligible overhead and unreliable aggregation in a mobile wireless network. In this study, we address this issue with a new \ac{OAC} method.

In the literature, \ac{FEEL} is investigated with several notable \ac{OAC} schemes. In \cite{Guangxu_2020}, the transmission of the local model parameters at the \acp{ED} over \ac{OFDM} subcarriers are proposed to achieve model parameter aggregation. To reverse the effect of the multipath channel on the transmitted signals, \ac{TCI} is applied, where the symbols on the \ac{OFDM} subcarriers are multiplied with the inverse of the channel coefficients and the subcarriers that fade are excluded from the transmissions. 
In \cite{Guangxu_2021}, \ac{OBDA}, inspired by distributed training by \ac{MV}~\cite{Bernstein_2018}, is proposed. In this method, the \acp{ED} transmit \ac{QPSK} symbols  over \ac{OFDM} subcarriers with \ac{TCI}, where the signs of the elements, i.e., votes, of the local stochastic gradient vectors to form the real and imaginary parts of the \ac{QPSK} symbols. At the \ac{ES}, the signs of the real and imaginary components of the superposed symbols on each subcarrier, i.e., the \ac{MV}, are used to estimate the global gradients.
%the majority of $1$s or $-1$s in the \ac{UL} determines the estimate of the global gradient vector. 
Despite the fact that \ac{OBDA} is compatible with digital modulations, for \ac{OAC}, each \ac{ED} still requires \ac{CSI} for \ac{TCI}. 
%In \cite{Amiri_2020}, each \acp{ED} projects gradient estimates into a sparse vector, resulting in a low-dimension vector for bandwidth reduction; the compressed data is then transmitted with \ac{BAA}. 
In \cite{Yang_2020} and \cite{Amiria_2021}, the \ac{CSI} is not available at the \acp{ED}, i.e., blind \acp{ED}. However, it is assumed that \ac{CSI} between each \ac{ED} and \ac{ES} is available at the \ac{ES}. %It is shown that beamforming with a large number of antennas can reduce the impact of the channel on the aggregation. 
To the best of our knowledge, there is no \ac{OAC} scheme where \ac{CSI} is unavailable to both the \acp{ED} and the \ac{ES} for \ac{FEEL} in the documented literature.

In this study, we propose an \ac{OAC} scheme for \ac{FEEL} without \ac{CSI} at the \acp{ED} and \ac{ES}. By considering distributed training by  \ac{MV}, we use \ac{PPM} to encode the votes, where the pulses are synthesized with \ac{DFT-s-OFDM}\cite{Sahin_2016}.
Since the proposed scheme determines the \ac{MV} with an energy detector applied to the superposed \ac{PPM} symbols, the \ac{CSI} is not needed at the \acp{ED} and the \ac{ES}. We also discuss the design with the consideration of the delay spread and the synchronization errors in the time domain. Finally, we prove the convergence of distributed training when the \ac{MV} is obtained with the proposed scheme under fading channel.

{\em Notation:} The complex and real numbers are denoted by $\complexNumbers$ and $\realNumbers$, respectively. 
$\expectationOperator[\cdot][]$ is the expectation. The sign function is denoted by $\sign(\cdot)$ and results in $1$, $-1$, or $\pm1$ at random for a positive, a negative, or a zero-valued argument, respectively. We use the notation  $(\textbf{a})_i^j$ as shorthand for denoting a vector $[{a}_i,{a}_{i+1},\mydots,{a}_j]^{\rm T}$. The $N$-dimensional all zero vector and $N\times N$ identity matrix are  $\zeroVector[{N}]$ and $\identityMatrix[{N}]$, respectively. $\indicatorFunction[\cdot]$ is the indicator function and $\probability[\cdot]$ is the probability of its argument.

\section{System Model}
\label{sec:system}

%With \ac{FEEL}, our main goal is to solve the same problem given in \eqref{eq:clp} without uploading the local data to the \ac{ES}.

\subsection{Distributed Training by Majority Vote}
Consider a network that consist of $\numberOfEdgeDevices$ \acp{ED} communicating with an \ac{ES}. Let $\dataset[\indexED]$ denote the local data containing labeled data samples at the $\indexED$th \ac{ED} as $\{(\sampleData[\indexSampleData], \sampleLabel[\indexSampleData] )\}\in\dataset[\indexED]$ for $\indexED=1,\mydots,\numberOfEdgeDevices$, where $\sampleData[\indexSampleData]$ and $\sampleLabel[\indexSampleData]$ are $\indexSampleData$th data sample and its associated label, respectively. In the case of availability of the local data at the \ac{ES} (e.g., each \ac{ED} uploads its data to the \ac{ES}), the centralized learning problem can be defined as 
\begin{align}
	\modelParametersOptimal=\arg\min_{\modelParameters} \lossFunctionGlobal[\modelParameters]=\arg\min_{\modelParameters} \frac{1}{|\completeData|}\sum_{\forall(\sampleData[], \sampleLabel[] )\in\completeData} \lossFunctionSample[{\modelParameters,\sampleData[],\sampleLabel[]}]~,
	\label{eq:clp}
\end{align}
where $\completeData=\dataset[1]\cup\dataset[2]\cup\cdots\cup\dataset[K]$ and  $\lossFunctionSample[{\modelParameters,\sampleData[],\sampleLabel[]}]$ is the sample loss function that measures the labeling error for $(\sampleData[], \sampleLabel[])$ for the parameters $\modelParameters=[\modelParametersEle[1],\mydots,\modelParametersEle[\numberOfModelParameters]]^{\rm T}\in\realNumbers^{\numberOfModelParameters}$, and $\numberOfModelParameters$ is the number of parameters. 
In the case of distributed training (e.g., the data is not available at the \ac{ES} as in \ac{FL}), one way of solving \eqref{eq:clp} relies on communicating the gradients between  the  \ac{ES} and  \acp{ED}. To reduce the cost of communicating gradients between the \ac{ES} and \acp{ED}, \ac{signSGD} with \ac{MV} is proposed in \cite{Bernstein_2018}. With this method, 
the updates at the $\indexCommunicationRound$th communication round can be expressed as
\begin{align}
\modelParametersAtIteration[\indexCommunicationRound+1] = \modelParametersAtIteration[\indexCommunicationRound] - \learningRate  \majorityVote[\indexCommunicationRound]~,
\label{eq:MVupdate}
\end{align}
where $\learningRate$ is the learning rate and $\majorityVote[\indexCommunicationRound]=[\majorityVoteEle[\indexCommunicationRound][1],\mydots,\majorityVoteEle[\indexCommunicationRound][\numberOfModelParameters]]^{\rm T}$ is the vector that contains the \acp{MV}. 
Assuming that  $|\dataset[\indexED]|=\cardinalityOfLocalData$ for $\indexED=1,\mydots,\numberOfEdgeDevices$, the $\indexGradient$th coordinate of $\majorityVote[\indexCommunicationRound]$ is calculated as
\begin{align}
\majorityVoteEle[\indexCommunicationRound][\indexGradient]\triangleq\sign\left(\sum_{\indexED=1}^{\numberOfEdgeDevices} {\signNormal[{\localGradientElement[\indexED,\indexGradient][\indexCommunicationRound]}]}\right)=\sign\left(\sum_{\indexED=1}^{\numberOfEdgeDevices} {\localGradientSignElement[\indexED,\indexGradient][\indexCommunicationRound]}\right)~,
\label{eq:majorityVote}
\end{align}
where $\localGradientElement[\indexED,\indexGradient][\indexCommunicationRound]$ is the $\indexGradient$th element of the local stochastic gradient vector given by
\begin{align}
\localGradient[\indexED][\indexCommunicationRound] =  \nabla \lossFunctionLocal[\indexED][{\modelParametersAtIteration[\indexCommunicationRound]}] 
= \frac{1}{\batchSize} \sum_{\forall(\sampleData[\indexSampleData], \sampleLabel[\indexSampleData] )\in\datasetBatch[\indexED]} \nabla 
\lossFunctionSample[{\modelParametersAtIteration[\indexCommunicationRound],\sampleData[\indexSampleData],\sampleLabel[\indexSampleData]}]
~,
	\label{eq:LocalGradientEstimate}
\end{align}
where $\datasetBatch[\indexED]\subset\dataset[\indexED]$ is the selected data batch from the local data samples and $\batchSize=|\datasetBatch[\indexED]|$ is the batch size. The corresponding \ac{MV}-based training procedure can be outlined as follows: The \ac{ES} first pulls $\localGradientSign[\indexED][\indexCommunicationRound]=[\localGradientSignElement[\indexED,1][\indexCommunicationRound],\mydots,\localGradientSignElement[\indexED,\numberOfModelParameters][\indexCommunicationRound]]^{\rm T}$ from all \acp{ED}. After calculating \eqref{eq:majorityVote}, $\forall\indexGradient$, it pushes the \ac{MV} vector $\majorityVote[\indexCommunicationRound]$ to the \acp{ED}. The \acp{ED} then update their parameters as in \eqref{eq:MVupdate} for the next communication round. Since this method communicates only the signs between \acp{ED} and \ac{ES}, it reduces the cost of communicating the gradients.
In this study, we consider the same training procedure for \ac{FEEL}. However, we use it over  a wireless network and develop an \ac{OAC} scheme to obtain the \ac{MV}, inspired by  \eqref{eq:majorityVote},  under fading channels.

%Therefore, the estimate of the global gradient can be calculated over the summations of the signs of local gradients given by

\subsection{Signal Model}
Consider a wireless network where each \ac{ED} and the \ac{ES}  are equipped with single antennas. We assume that the  \acp{ED}' average signal powers are identical at the \ac{ES}'s location (i.e., large-scale impacts of the channel are compensated) and managed with an uplink power control mechanism, e.g., through \ac{PRACH} and/or \ac{PUCCH} in 3GPP \ac{5G} \ac{NR} \cite{10.5555/3294673}.  We consider the fact the time-synchronization  between the \acp{ED} may not be perfect and the maximum difference between time of arriving \acp{ED} signals at the \ac{ES}'s location is $\syncError$~seconds.

We assume that the \acp{ED} access the wireless channel  on the same time-frequency resources with $\numberOfOFDMSymbols$ \ac{DFT-s-OFDM}  symbols at the $\indexCommunicationRound$th  round. The $\indexOFDMSymbol$th transmitted baseband \ac{DFT-s-OFDM} symbol in discrete 
time for the $\indexED$th \ac{ED} can be expressed as 
\begin{align}
    \transmittedVector[\indexED,\indexOFDMSymbol] = \idftMatrix[\idftSize]\frequencyMapping\transformPrecoder[\numberOfActiveSubcarriers]\symbolVector[\indexED,\indexOFDMSymbol]~,
\end{align}
where $\idftMatrix[\idftSize]\in\complexNumbers^{\idftSize\times\idftSize}$ is the $\idftSize$-point \ac{IDFT} matrix, $\transformPrecoder[\numberOfActiveSubcarriers]\in\complexNumbers^{\numberOfActiveSubcarriers\times\numberOfActiveSubcarriers}$ is the $\numberOfActiveSubcarriers$-point \ac{DFT} matrix,  $\frequencyMapping\in\realNumbers^{\idftSize\times\numberOfActiveSubcarriers}$ is the mapping matrix that maps the output of the \ac{DFT} precoder to a set of contiguous subcarriers,
%$\CPadder\in\realNumbers^{\idftSize+\cpSize\times \idftSize}$ is a matrix that prepends \ac{CP} of length $\cpSize$,
and $\symbolVector[\indexED,\indexOFDMSymbol]\in\complexNumbers^{\numberOfActiveSubcarriers}$ contains the symbols on $\numberOfActiveSubcarriers$ bins. Note that  \ac{DFT-s-OFDM} is a special \ac{SC} waveform using circular convolution \cite{Sahin_2016}, where the symbol spacing in time is $\symbolSpacing=\idftSize\samplePeriod/\numberOfActiveSubcarriers$~seconds, the pulse shape is Dirichlet sinc \cite{Kakkavas_2017}, and $\samplePeriod$ is the sample period.

In this study, we assume the \ac{CP} duration is larger than the maximum-excess delay denoted by $\channelSpread$~seconds. Hence, assuming the transmissions from the \acp{ED} arrive at the \ac{ES} within the \ac{CP} duration, the  $\indexOFDMSymbol$th received baseband signal in discrete-time can be written as
\begin{align}
    \receivedVector[\indexOFDMSymbol] =\sum_{\indexED=1}^{\numberOfEdgeDevices}\channelMatrix[\indexED]\transmittedVector[\indexED,\indexOFDMSymbol]+\noiseVector[\indexOFDMSymbol]~,
\end{align}
where $\channelMatrix[\indexED]\in\complexNumbers^{\idftSize\times\idftSize}$ is a circular-convolution matrix based on the \ac{CIR} between the $\indexED$th \ac{ED} and the \ac{ES} and $\noiseVector[\indexOFDMSymbol]\sim\mathcal{CN}(\zeroVector[{\idftSize}],\noiseVariance\identityMatrix[{\idftSize}])$ is the \ac{AWGN}. At the \ac{ES}, we calculate the aggregated symbols on the bins as
$\symbolVectorEstimate[\indexOFDMSymbol] =\transformDecoder[\numberOfActiveSubcarriers]\frequencyMapping^{\rm H}\dftMatrix[\idftSize]\receivedVector[\indexOFDMSymbol]
$. Note that we do not use \ac{FDE} since we use \ac{DFT-s-OFDM}  for calculating the \ac{MV} with a non-coherent detector.

\subsection{Performance Metrics}
\subsubsection{PMEPR} We define the \ac{PMEPR} as  $\max_{\timeVar\in[0, \symbolDuration)}|\timeDomainOFDM[\timeVar]|^2/{\Ptransmit}
$, where  $\timeDomainOFDM[\timeVar]\in \complexNumbers$ is the baseband \ac{OFDM}/\ac{DFT-s-OFDM} symbol in continuous time, $\symbolDuration$ is the symbol duration, and $\Ptransmit=\expectationOperator[{|\timeDomainOFDM[\timeVar]|^2}][\timeVar]=\numberOfActiveSubcarriers/\idftSize$ is the mean-envelope power as $\norm{{\symbolVector[\indexED,\indexOFDMSymbol]}}_2^2$ is equal to $\numberOfActiveSubcarriers$ since all bins are actively utilized.
\subsubsection{Convergence rate}
In this study, we define  the convergence rate \cite{Guangxu_2021,Bernstein_2018} as the rate at which the expected value of average norm of the gradient of $ \lossFunctionGlobal[\modelParameters]$  diminishes  as the number of total communication rounds $\communicationRounds$ and $\numberOfEdgeDevices$ when the training is done in the presence of the proposed scheme.

%\subsubsection{Test accuracy}
%Test accuracy is defined as how good model the perform the classification for unobserved data samples.

\section{Majority Vote with PPM in Fading Channel}
\label{sec:ppmMV}
\subsection{Edge Devices - Transmitter}
At the $\indexED$th \ac{ED}'s transmitter, we encode the signs of the local gradients, i.e., $\{\localGradientSignElement[\indexED,\indexGradient][\indexCommunicationRound]\}$, $\forall\indexGradient,\indexED$, with \ac{PPM}. We synthesize the pulse in a \ac{PPM} symbol by activating consecutive $\numberOfElementsInAPulse$ bins of \ac{DFT-s-OFDM}, which effectively corresponds to a pulse with the duration of $\pulseDuration\approx\numberOfElementsInAPulse\symbolSpacing$~seconds by combining $\numberOfElementsInAPulse$ shifted versions of the Dirichlet sinc functions in time. To accommodate the time-synchronization errors  and the delay spread, we consider a guard period  between the adjacent pulses. Thus, we deactivate the following $\uniformGap$ bins after $\numberOfElementsInAPulse$ active bins, which results in a guard period with the duration of $\guardTimeUniform\approx\uniformGap\symbolSpacing$~seconds, where $\guardTimeUniform\ge\channelSpread+\syncError$ must hold true. As a result, the maximum number of votes that can be carried for each \ac{DFT-s-OFDM} symbol can be calculated as
%\begin{align}
$    \numberOfVotesPerDFTsOFDM = \floor*{\frac{\numberOfActiveSubcarriers}{2(\numberOfElementsInAPulse+\uniformGap)}}$, 
%\end{align}
where $\uniformGap\ge\ceil{(\channelSpread+\syncError)/\symbolSpacing}$.

In this study, we consider a generalized mapping rule that maps the signs of the local gradients to the positions of the pulses within a \ac{DFT-s-OFDM} symbol and $\numberOfOFDMSymbols$ \ac{DFT-s-OFDM} symbols. To this end, let $\mappingFunction$ be a function that maps $\indexGradient\in\{1,2,\mydots,\numberOfModelParameters\}$ to the distinct pairs $(\voteInTime[+],\voteInFrequency[+])$ and $(\voteInTime[-],\voteInFrequency[-])$ that indicate the pulse positions for $\voteInTime[+],\voteInTime[-]\in\{0,1,\mydots,\numberOfOFDMSymbols-1\}$ and $\voteInFrequency[+],\voteInFrequency[-]\in\{0,1,\mydots,2\numberOfVotesPerDFTsOFDM-1\}$. For all $\indexGradient$, we then determine the following bins of \ac{DFT-s-OFDM} symbols as
\begin{align}
	(\symbolVector[\indexED,{\voteInTime[+]}])_{\voteInFrequency[+](\numberOfElementsInAPulse+\uniformGap)}^{\voteInFrequency[+](\numberOfElementsInAPulse+\uniformGap)+\numberOfElementsInAPulse-1} =  \symbolsActivatedBins \randomSymbolAtSubcarrier[\indexED,\indexGradient]\indicatorFunction[{\localGradientSignElement[\indexED,\indexGradient][\indexCommunicationRound] =1}]~,
	\nonumber
\end{align}
and
\begin{align}
	(\symbolVector[\indexED,{\voteInTime[-]}])_{\voteInFrequency[-](\numberOfElementsInAPulse+\uniformGap)}^{\voteInFrequency[-](\numberOfElementsInAPulse+\uniformGap)+\numberOfElementsInAPulse-1} = \symbolsActivatedBins \randomSymbolAtSubcarrier[\indexED,\indexGradient]\indicatorFunction[{\localGradientSignElement[\indexED,\indexGradient][\indexCommunicationRound] =-1}]~,
	\nonumber
\end{align}
%Define $\symbolVectorNotation_{\voteInTime[+],\voteInFrequency[+]},\symbolVectorNotation_{\voteInTime[-],\voteInFrequency[-]}\in\complexNumbers^{\numberOfElementsInAPulse}$  as
%\begin{align}
%	\symbolVectorNotation_{\voteInTime[+],\voteInFrequency[+]}\triangleq \symbolsActivatedBins \randomSymbolAtSubcarrier[\indexED,\indexGradient]\indicatorFunction[{\localGradientSignElement[\indexED,\indexGradient][\indexCommunicationRound] =1}]~,
%\label{eq:symbolOne}
%\end{align}
%and
%\begin{align}
%	\symbolVectorNotation_{\voteInTime[-],\voteInFrequency[-]}\triangleq\symbolsActivatedBins \randomSymbolAtSubcarrier[\indexED,\indexGradient]\indicatorFunction[{\localGradientSignElement[\indexED,\indexGradient][\indexCommunicationRound] =-1}]~,
%\label{eq:symbolTwo}
%\end{align}
where $\symbolsActivatedBins\in\complexNumbers^{\numberOfElementsInAPulse}$ contains the weights of the Dirichlet sinc functions to generate the pulse, and $\randomSymbolAtSubcarrier[\indexED,\indexGradient]$ is a random symbol on the unit-circle. 
%We then map $\symbolVectorNotation_{\voteInTime[+],\voteInFrequency[+]}$ and $\symbolVectorNotation_{\voteInTime[-],\voteInFrequency[-]}$ to the bins of \ac{DFT-s-OFDM} symbols as
%\begin{align}
%   (\symbolVector[\indexED,{\voteInTime[+]}])_{\voteInFrequency[+](\numberOfElementsInAPulse+\uniformGap)}^{\voteInFrequency[+](\numberOfElementsInAPulse+\uniformGap)+\numberOfElementsInAPulse-1} = \symbolVectorNotation_{\voteInTime[+],\voteInFrequency[+]}
%\end{align}
%and
%\begin{align}
%   (\symbolVector[\indexED,{\voteInTime[-]}])_{\voteInFrequency[-](\numberOfElementsInAPulse+\uniformGap)}^{\voteInFrequency[-](\numberOfElementsInAPulse+\uniformGap)+\numberOfElementsInAPulse-1} = \symbolVectorNotation_{\voteInTime[-],\voteInFrequency[-]}~,
%\end{align}
%respectively. 
Therefore, the proposed scheme defines two pulse positions over two different time resources for one vote. If $\voteInTime[+]=\voteInTime[-]$ and $\voteInFrequency[+]=\voteInFrequency[-]+1$ for all $\indexGradient$, the adjacent time resources of $\voteInTime[+]$th \ac{DFT-s-OFDM} symbol are used for voting. The \ac{MV} calculation with proposed scheme under this specific mapping is referred to as \ac{PPM-MV} in this study. %Without loss of generality, we assume that \ac{GM} keeps the gradient orders.

\subsubsection{Pulse Shape}
We choose $\symbolsActivatedBins$ as $\sqrt{\symbolEnergy}\times[1,-1,1,-1,\cdots]^{\rm T}$ since this sequence yields a rectangular-like pulse shape in the time domain for \ac{DFT-s-OFDM}, as illustrated in Section~\ref{sec:numerical}, where $\symbolEnergy=2(\numberOfElementsInAPulse+\uniformGap)/\numberOfElementsInAPulse$ is an energy normalization factor. It is worth noting that the proposed framework allows one to design $\symbolsActivatedBins$ for various pulse shapes, which can be considered for further optimization of the proposed scheme. 

\begin{figure*}[t]
	\centering
	{\includegraphics[width =6.5in]{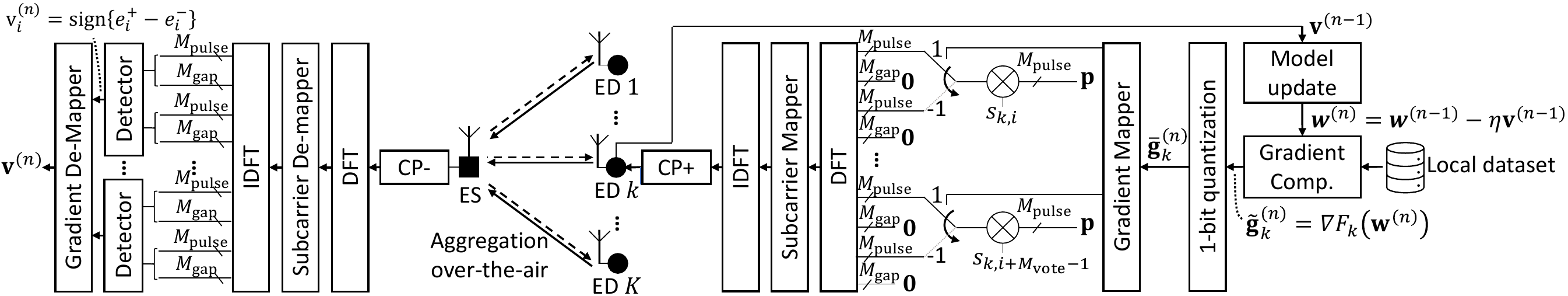}
	} 
	\caption{Transmitter and receiver diagrams for \ac{PPM-MV} with \ac{DFT-s-OFDM} for \ac{FEEL}.}
	\label{fig:feelBlockDiagram}
	\vspace{-2mm}
\end{figure*}
%\begin{figure}
%	\centering
%	\subfloat[ED1's votes.]{\includegraphics[width =\figuresizeA]{plot_exampleA.pdf}
%		\label{subfig:a}}\vspace{-3mm}\\
%	\subfloat[ED2's votes.]{\includegraphics[width =\figuresizeA]{plot_exampleB.pdf}
%		\label{subfig:b}}\vspace{-3mm}\\
%	\subfloat[ED3's votes.]{\includegraphics[width =\figuresizeA]{plot_exampleC.pdf}
%		\label{subfig:c}}\vspace{-3mm}\\	
%	\subfloat[ED1's signal. A pulse is a combination of Dirichlet sinc functions.]{\includegraphics[width =\figuresizeA]{plot_exampleD.pdf}
%		\label{subfig:d}}\vspace{-3mm}	\\
%	\subfloat[ED1's signal including time dispersion and  synchronization error.]{\includegraphics[width =\figuresizeA]{plot_exampleE.pdf}
%		\label{subfig:e}}\vspace{-3mm}\\
%	\subfloat[ED1's vote at the ES location.]{\includegraphics[width =\figuresizeA]{plot_exampleF.pdf}
%		\label{subfig:f}}\vspace{-3mm}\\	
%	\subfloat[Aggregated votes over the air and the energy detection at the ES.]{\includegraphics[width =\figuresizeA]{plot_exampleG.pdf}
%		\label{subfig:g}}			
%	\caption{An example of \ac{MV} with \ac{PPM-MV} based on \ac{DFT-s-OFDM} with $\numberOfEdgeDevices=3$ \acp{ED} ($\numberOfElementsInAPulse=4,\uniformGap=3,\numberOfVotesPerDFTsOFDM=3,\numberOfActiveSubcarriers=42, \idftSize=2048)$}
%	\label{fig:conceptBins}
%	\vspace{-2mm}
%\end{figure}

\subsection{Edge Server - Receiver}
At the ES, we first calculate the pairs $(\voteInTime[+],\voteInFrequency[+])$ and $(\voteInTime[-],\voteInFrequency[-])$  based on $\mappingFunction$ for a given $\indexGradient$. We then obtain the \ac{MV}  for the $\indexGradient$th gradient with an energy detector as
\begin{align}
	\majorityVoteEle[\indexCommunicationRound][\indexGradient] = \signNormal[{\deltaVectorAtIteration[\indexCommunicationRound][\indexGradient]}]~,
	\label{eq:detector}
\end{align}
where $\deltaVectorAtIteration[\indexCommunicationRound][\indexGradient]\triangleq{\metricForFirst[\indexGradient]-\metricForSecond[\indexGradient]}$ for
\begin{align}
	\metricForFirst[\indexGradient]\triangleq  \norm{(\symbolVectorEstimate[{\voteInTime[+]}])_{\voteInFrequency[+](\numberOfElementsInAPulse+\uniformGap)}^{\voteInFrequency[+](\numberOfElementsInAPulse+\uniformGap)+\numberOfElementsInAPulse+\uniformGap-1}}_2^2~,
	\label{eq:energyFirst}
\end{align}
and
\begin{align}
	\metricForSecond[\indexGradient] \triangleq  \norm{(\symbolVectorEstimate[{\voteInTime[-]}])_{\voteInFrequency[-](\numberOfElementsInAPulse+\uniformGap)}^{\voteInFrequency[-](\numberOfElementsInAPulse+\uniformGap)+\numberOfElementsInAPulse+\uniformGap-1}}_2^2~.
	\label{eq:energySecond}
\end{align}
Since the multipath channel disperses the pulses in the time domain and the synchronization error changes the position of the pulse in time, we consider $\numberOfElementsInAPulse+\uniformGap$ bins for the energy calculations in \eqref{eq:energyFirst} and \eqref{eq:energySecond}.

In \figurename~\ref{fig:feelBlockDiagram}, the transmitter and the receiver block diagrams are provided based on the aforementioned discussions. 
%The \ac{MV} principle with the proposed scheme is also exemplified in Figure~\ref{fig:conceptBins} for $\numberOfEdgeDevices=3$ \acp{ED}, $\numberOfElementsInAPulse=4$, $\uniformGap=3$, $\numberOfVotesPerDFTsOFDM=3$, $\numberOfActiveSubcarriers=42$, and $\idftSize=2048$. In Figure~\ref{fig:conceptBins}\subref{subfig:a}-\subref{subfig:c}, the votes from three \acp{ED} are provided for three gradients. In Figure~\ref{fig:conceptBins}\subref{subfig:d}, the first \ac{ED}'s signal in the time domain is shown, where the pulses are generated based on the summation of shifted Dirichlet sinc functions. In  Figure~\ref{fig:conceptBins}\subref{subfig:e}, the first \ac{ED}'s signal includes the synchronization error and the time dispersion due to the multi-path channel.  In  Figure~\ref{fig:conceptBins}\subref{subfig:f}, the \ac{ED}'s votes (excluding the other \acp{ED}) after the receiver processing is shown. In  Figure~\ref{fig:conceptBins}\subref{subfig:g}, the aggregated votes from all \acp{ED} on the bins are shown. Based on this example, the \ac{MV} vector at the \ac{ES} can be calculated as  $\majorityVote[\indexCommunicationRound]=(\majorityVoteEle[\indexCommunicationRound][0],\majorityVoteEle[\indexCommunicationRound][1],\majorityVoteEle[\indexCommunicationRound][2])=(1,0,-1)$ based on \eqref{eq:detector}, where we assume a tie for the second gradient.  It is worth nothing that the probability of tie is zero in practice due to the \ac{AWGN} at the \ac{ES}' receiver.

\subsection{Why Does It Work without CSI at the EDs and ES?}
\label{ssec:why}
Let $\numberOFEDsForOptionOne$ and $\numberOFEDsForOptionSecond=\numberOfEdgeDevices-\numberOFEDsForOptionOne$ be the numbers of \acp{ED} that contribute a vote towards $1$ and $-1$ for the $\indexGradient$th gradient, respectively. It is trivial to show that ${\metricForFirst[\indexGradient]}$ and ${\metricForSecond[\indexGradient]}$ are exponential random variables, where their means are approximately
$\meanOptionOne\triangleq\expectationOperator[{\metricForFirst[\indexGradient]}][]\approx\numberOfElementsInAPulse\symbolEnergy\numberOFEDsForOptionOne+(\numberOfElementsInAPulse + \uniformGap)\noiseVariance$ and $\meanOptionTwo\triangleq\expectationOperator[{\metricForSecond[\indexGradient]}][]
\approx\numberOfElementsInAPulse\symbolEnergy\numberOFEDsForOptionSecond+(\numberOfElementsInAPulse + \uniformGap)\noiseVariance$.\footnote{The  reason for the approximation is that the interference between \ac{PPM} symbols in a multipath channel  is assumed to be negligible for a large $\uniformGap$.} Since $\meanOptionOne$ and $\meanOptionTwo$ are linear functions of $\numberOFEDsForOptionOne$ and $\numberOFEDsForOptionSecond$, respectively, the proposed scheme obtains the correct  \ac{MV} {\em probabilistically} as the \ac{PPM} symbols may not coherently add up and their amplitudes may not be aligned in fading channel. Therefore, the \ac{MV} calculated in \eqref{eq:detector} is different from the original \ac{MV} given in \eqref{eq:majorityVote}. Hence, to provide a convincing answer to the question  if the proposed scheme maintains the convergence of the original \ac{MV} in \cite{Bernstein_2018}, we need to show the convergence for a non-convex loss function $\lossFunctionGlobal[\modelParameters]$. To this end, we consider several standard assumptions made in the literature \cite{Bernstein_2018, Guangxu_2021}:
\begin{assumption}[Bounded loss function]
	\rm 
	$\lossFunctionGlobal[\modelParameters]\ge \lossFunctionGlobalMinimum$, $\forall\modelParameters$. 
\end{assumption}
\begin{assumption}[Smooth] 
	\rm 
	Let $\globalGradientNoIndex$ be the gradient of $\lossFunctionGlobal[\modelParameters]$ evaluated at $\modelParameters$. For all $\modelParameters$ and $\modelParameters'$, the expression given by
	\begin{align}
		\left| \lossFunctionGlobal[\modelParameters'] - (\lossFunctionGlobal[\modelParameters]+\globalGradientNoIndex^{\rm T}(\modelParameters'-\modelParameters)) \right| \le \frac{1}{2}\sum_{\indexGradient=1}^{\numberOfModelParameters} \nonnegativeConstantsEle[\indexGradient](\modelParametersEle[\indexGradient]'-\modelParametersEle[\indexGradient])^2~,
		\nonumber
	\end{align}	
	holds for a non-negative constant vector
	$\nonnegativeConstants=[\nonnegativeConstantsEle[1],\mydots,\nonnegativeConstantsEle[\numberOfModelParameters]]^{\rm T}$.
\end{assumption}
\begin{assumption}[Variance bound]
	\rm The stochastic  gradient estimates $\{\localGradientNoIndex[\indexED]=[\localGradientNoIndexElement[\indexED,1],\mydots,\localGradientNoIndexElement[\indexED,\numberOfModelParameters]]^{\rm T}=\nabla \lossFunctionLocal[\indexED][{\modelParametersAtIteration[\indexCommunicationRound]}]\} $, $\forall\indexED$, are independent and  unbiased estimates of $\globalGradientNoIndex=[\globalGradientElementNoIndex[1],\mydots,\globalGradientElementNoIndex[\numberOfModelParameters]]^{\rm T}=\nabla\lossFunctionGlobal[{\modelParameters}]$ with a coordinate bounded variance, i.e.,
	\begin{align}
		\expectationOperator[{\localGradientNoIndex[\indexED]}][]&=\globalGradientNoIndex,~\forall\indexED,\\	\expectationOperator[{(\localGradientNoIndexElement[\indexED,\indexGradient]-\globalGradientElementNoIndex[\indexGradient])^2}][]&\le\varianceBoundEle[\indexGradient]^2/\batchSize,~\forall\indexED,\indexGradient,
	\end{align}
	where  $\varianceBound = [\varianceBoundEle[1],\mydots,\varianceBoundEle[\numberOfModelParameters]]^{\rm T}$ is a non-negative constant  vector.
\end{assumption}
\begin{assumption}[Unimodal, symmetric gradient noise]
	\rm
	For any given $\modelParameters$, the elements of the vector $\localGradientNoIndex[\indexED]$, $\forall\indexED$, has a unimodal distribution that is also symmetric around its mean.
\end{assumption}

\begin{theorem}
\rm For $\batchSize=\communicationRounds/\brachSizeRelativeToRounds$ and $\learningRate=1/\sqrt{\norm{\nonnegativeConstants}_1\batchSize}$, the convergence rate of the distributed training by the \ac{MV} based on \ac{PPM} in fading channel is
\begin{align}
\expectationOperator[\frac{1}{\communicationRounds}\sum_{\indexCommunicationRound=0}^{\communicationRounds-1} \norm{\globalGradient[\indexCommunicationRound]}_1][]\nonumber\le\frac{1}{\sqrt{\communicationRounds}}&\bigg( \coefficientOne\sqrt{\norm{\nonnegativeConstants}_1}\left(	\lossFunctionGlobal[{\modelParametersAtIteration[0]}]- \lossFunctionGlobalMinimum+\frac{\brachSizeRelativeToRounds}{2}\right)\nonumber\\&~~+\frac{2\sqrt{2\brachSizeRelativeToRounds}}{3}\norm{\varianceBound}_1\bigg)~,
\label{eq:convergence}
\end{align}
where  $\coefficientOne=(1+\frac{2}{\effectiveSNR\numberOfEdgeDevices})\frac{1}{\sqrt{\brachSizeRelativeToRounds}}$  for $\effectiveSNR\triangleq\frac{\numberOfElementsInAPulse\symbolEnergy}{(\numberOfElementsInAPulse + \uniformGap)\noiseVariance} $.
\label{th:convergence}
\end{theorem}
\begin{IEEEproof}
	%The proof of Theorem~\ref{th:convergence} relies on a well-known strategy of relating the norm of the gradient of the loss function $ \lossFunctionGlobal[\modelParameters]$ to the expected improvement made in a single step as described in \cite{Bernstein_2018}.
	 Let $\globalGradient[\indexCommunicationRound]$ be the gradient of $\lossFunctionGlobal[{\modelParametersAtIteration[\indexCommunicationRound]}]$ (i.e., the true gradient                                                                                                                                                                                                                                                                                                                                                                                                                                                                                                                                                                                                                                                     ). By using Assumption~2 and using \eqref{eq:detector}, we can write
	\begin{align}
		\lossFunctionGlobal[{\modelParametersAtIteration[\indexCommunicationRound+1]}]& - \lossFunctionGlobal[{\modelParametersAtIteration[\indexCommunicationRound]}]\le -\learningRate{\globalGradient[\indexCommunicationRound]}^{\rm T}\majorityVote[\indexCommunicationRound] + \frac{\learningRate^2}{2}\norm{\nonnegativeConstants}_1\nonumber\\
		=&-\learningRate\norm{\globalGradient[\indexCommunicationRound]}_1+\frac{\learningRate^2}{2}\norm{\nonnegativeConstants}_1\nonumber\\&+2\learningRate\sum_{\indexGradient=1}^{\numberOfModelParameters}|\globalGradientElement[\indexCommunicationRound][\indexGradient]| \indicatorFunction[{\signNormal[{\deltaVectorAtIteration[\indexCommunicationRound][\indexGradient]}]\neq \signNormal[{\globalGradientElement[\indexCommunicationRound][\indexGradient]}]}]\nonumber~.
	\end{align}	
	Thus,
	\begin{align}
		&\expectationOperator[{	\lossFunctionGlobal[{\modelParametersAtIteration[\indexCommunicationRound+1]}] - \lossFunctionGlobal[{\modelParametersAtIteration[\indexCommunicationRound]}]|\modelParametersAtIteration[\indexCommunicationRound]}][] \le  -\learningRate\norm{\globalGradient[\indexCommunicationRound]}_1+\frac{\learningRate^2}{2}\norm{\nonnegativeConstants}_1\nonumber\\&~~~~~~~~~~~~~~+\underbrace{2\learningRate\sum_{\indexGradient=1}^{\numberOfModelParameters}|\globalGradientElement[\indexCommunicationRound][\indexGradient]| \underbrace{\probability[{\signNormal[{\deltaVectorAtIteration[\indexCommunicationRound][\indexGradient]}]\neq \signNormal[{\globalGradientElement[\indexCommunicationRound][\indexGradient]}]}]\nonumber}_{\triangleq\probabilityIncorrect[\indexGradient]}}_{\text{Stochasticity-induced error}}~.
	\end{align}
	A bound on the stochasticity-induced error can be obtained as follows: Assume that $\signNormal[{\globalGradientElement[\indexCommunicationRound][\indexGradient]}]=1$. Let $\numberOfEDsWithCorrectChoice$ be a random variable for counting the number of EDs with the correct decision, i.e., $\signNormal[{\globalGradientElement[\indexCommunicationRound][\indexGradient]}]=1$. The random variable $\numberOfEDsWithCorrectChoice$ can  then be model as the sum of $\numberOfEdgeDevices$ independent Bernoulli trials, i.e., a binomial variable with the success and failure probabilities given by
	\begin{align}
		\correctDecision[\indexGradient]\triangleq\probability[{\signNormal[{\localGradientElement[\indexED,\indexGradient][\indexCommunicationRound]}]= \signNormal[{\globalGradientElement[\indexCommunicationRound][\indexGradient]}]}]\nonumber~,\\
		\incorrectDecision[\indexGradient]\triangleq\probability[{\signNormal[{\localGradientElement[\indexED,\indexGradient][\indexCommunicationRound]}]\neq \signNormal[{\globalGradientElement[\indexCommunicationRound][\indexGradient]}]}]\nonumber~,
	\end{align}
	respectively, for all $\indexED$. This implies that
	\begin{align}
		\probabilityIncorrect[\indexGradient]=
		\sum_{\numberOFEDsForOptionOne=0}^{\numberOfEdgeDevices}
		\probability[{\signNormal[{\deltaVectorAtIteration[\indexCommunicationRound][\indexGradient]}]\neq 1}|\numberOfEDsWithCorrectChoice=\numberOFEDsForOptionOne]\probability[\numberOfEDsWithCorrectChoice=\numberOFEDsForOptionOne]~, \nonumber
	\end{align}
	where
	%\begin{align}
		$\probability[\numberOfEDsWithCorrectChoice=\numberOFEDsForOptionOne] = \binom{\numberOfEdgeDevices}{\numberOFEDsForOptionOne}\correctDecision[\indexGradient]^{\numberOFEDsForOptionOne}\incorrectDecision[\indexGradient]^{\numberOfEdgeDevices-\numberOFEDsForOptionOne}$.
	%	\label{eq:binProb}
	%\end{align}
	To calculate $\probability[{\signNormal[{\deltaVectorAtIteration[\indexCommunicationRound][\indexGradient]}]\neq 1}|\numberOfEDsWithCorrectChoice=\numberOFEDsForOptionOne]$, we use the distribution of $\deltaVectorAtIteration[\indexCommunicationRound][\indexGradient]$, which can be obtained by using the properties of exponential random variables as
	\begin{align}
		f(\deltaVectorAtIteration[\indexCommunicationRound][\indexGradient]) = \begin{cases} 
			\frac{\constante^{-\frac{\deltaVectorAtIteration[\indexCommunicationRound][\indexGradient]}{\meanOptionTwo}}}{\meanOptionOne+\meanOptionTwo}, & \deltaVectorAtIteration[\indexCommunicationRound][\indexGradient]\le 0 \\
			\frac{\constante^{-\frac{\deltaVectorAtIteration[\indexCommunicationRound][\indexGradient]}{\meanOptionOne}}}{\meanOptionOne+\meanOptionTwo}, & \deltaVectorAtIteration[\indexCommunicationRound][\indexGradient]>0 
		\end{cases}~.
		\label{eq:lappro}
	\end{align}
	Thus, by integrating \eqref{eq:lappro} with respect to $\deltaVectorAtIteration[\indexCommunicationRound][\indexGradient]$,
	\begin{align}
		\probability[{\signNormal[{\deltaVectorAtIteration[\indexCommunicationRound][\indexGradient]}]\neq 1}|\numberOfEDsWithCorrectChoice=\numberOFEDsForOptionOne]& = \frac{\meanOptionTwo}{\meanOptionOne+\meanOptionTwo}
		%\\&~~~~~~~~~=\frac{\numberOfElementsInAPulse\symbolEnergy(\numberOfEdgeDevices-\numberOFEDsForOptionOne)+(\numberOfElementsInAPulse + \uniformGap)\noiseVariance}{\numberOfElementsInAPulse\symbolEnergy\numberOfEdgeDevices+2(\numberOfElementsInAPulse + \uniformGap)\noiseVariance}\nonumber
		=\frac{(\numberOfEdgeDevices-\numberOFEDsForOptionOne)+1/\effectiveSNR}{\numberOfEdgeDevices+2/\effectiveSNR}.
		\label{eq:probLapResult}
	\end{align}
	Hence, by using \eqref{eq:probLapResult} and the properties of binomial coefficients
	\begin{align}
		\probabilityIncorrect[\indexGradient]&=\sum_{\numberOFEDsForOptionOne=0}^{\numberOfEdgeDevices}
		\frac{(\numberOfEdgeDevices-\numberOFEDsForOptionOne)+1/\effectiveSNR}{1+2/\effectiveSNR}\binom{\numberOfEdgeDevices}{\numberOFEDsForOptionOne}\correctDecision[\indexGradient]^{\numberOFEDsForOptionOne}\incorrectDecision[\indexGradient]^{\numberOfEdgeDevices-\numberOFEDsForOptionOne}=\frac{\frac{1}{\effectiveSNR\numberOfEdgeDevices}+\incorrectDecision[\indexGradient]}{1+\frac{2}{\numberOfEdgeDevices\effectiveSNR}}~.\nonumber
	\end{align}
	Under Assumption 2 and Assumption 3, by using the derivations in \cite{Bernstein_2018},  $\incorrectDecision[\indexGradient]\le\frac{\sqrt{2}\varianceBoundEle[\indexGradient]}{3|\globalGradientElement[\indexCommunicationRound][\indexGradient]|\sqrt{\batchSize}}$ holds true. Hence, an upper bound on the stochasticity-induced error can be obtained as
	\begin{align}
		\sum_{\indexGradient=1}^{\numberOfModelParameters}|\globalGradientElement[\indexCommunicationRound][\indexGradient]|\probabilityIncorrect[\indexGradient]\le  \frac{\frac{1}{\effectiveSNR\numberOfEdgeDevices}}{1+\frac{2}{\numberOfEdgeDevices\effectiveSNR}}\norm{\globalGradient[\indexCommunicationRound]}_1 +\frac{1}{\sqrt{\batchSize}} \frac{\sqrt{2}/3}{1+\frac{2}{\numberOfEdgeDevices\effectiveSNR}}\norm{\varianceBound}_1~.\nonumber
	\end{align}
	Based on Assumption~1, we can show that
	\begin{align}
		&\lossFunctionGlobal[{\modelParametersAtIteration[0]}]-\lossFunctionGlobalMinimum\ge \lossFunctionGlobal[{\modelParametersAtIteration[0]}]-\expectationOperator[{\lossFunctionGlobal[{\modelParametersAtIteration[\communicationRounds]}]}][]\nonumber\\&=\expectationOperator[{\sum_{\indexCommunicationRound=0}^{\communicationRounds-1}\lossFunctionGlobal[{\modelParametersAtIteration[\indexCommunicationRound]}] - \lossFunctionGlobal[{\modelParametersAtIteration[\indexCommunicationRound+1]}]}][]\nonumber\\
		&\ge
		\expectationOperator[{	\sum_{\indexCommunicationRound=0}^{\communicationRounds-1}\frac{\learningRate}{1+\frac{2}{\numberOfEdgeDevices\effectiveSNR}}\norm{\globalGradient[\indexCommunicationRound]}_1-\frac{\learningRate^2}{2}\norm{\nonnegativeConstants}_1 -\frac{\learningRate}{\sqrt{\batchSize}} \frac{2\sqrt{2}/3}{1+\frac{2}{\numberOfEdgeDevices\effectiveSNR}}\norm{\varianceBound}_1  }][]~.
		\label{eq:finaleq}
	\end{align}
	By rearranging the terms in \eqref{eq:finaleq} and using the expressions for $\batchSize$ and $\learningRate$, \eqref{eq:convergence} is reached.
\end{IEEEproof}

Theorem~\ref{th:convergence} shows that when $\effectiveSNR$  and $\numberOfEdgeDevices$ are large, the convergence with the proposed scheme in fading channel is similar to the one with \ac{signSGD} in an ideal channel \cite[Theorem~1]{Bernstein_2018}. 

\subsection{Implementation Details, Trade-offs, and Comparisons}
The main difference of the proposed scheme as compared to  the approaches in \cite{Guangxu_2020} and \cite{Guangxu_2021} is that it does not need \ac{TCI} at the \acp{ED} and prevents the loss of the gradients due to the truncation. As opposed to the methods in \cite{Yang_2020} and \cite{Amiria_2021}, it also does not require \ac{CSI} at the \ac{ES} or multiple antennas. Therefore, the proposed scheme offers practical distributed learning in mobile networks.
The second major difference of the proposed scheme is that it leads to an interesting trade-off between \ac{PMEPR} and resource utilization as shown in Section~\ref{sec:numerical}. For a given $\uniformGap$,  as $\numberOfElementsInAPulse$ increases, the pulse energy distributes more evenly in time and the amplitude decreases as less votes are carried. This results in a decreasing \ac{PMEPR}, but more resource consumption. %The third difference is that the randomization symbols in \eqref{eq:symbolOne} or \eqref{eq:symbolTwo} lead to a better accuracy results for non-\ac{iid} data. 
The shortcoming of the proposed scheme is that it consumes a larger number of \ac{DFT-s-OFDM} symbols as compared to \ac{OBDA}. Although this appears to be a limitation, we emphasize that the proposed method eliminates the non-negligible channel estimation overhead and is immune to the time-variations of the channel and time synchronization error. Finally, we choose $\randomSymbolAtSubcarrier[\indexED,\indexGradient]$ as random \ac{QPSK} in this study since this is implementation-friendly and randomizes ${\metricForFirst[\indexGradient]}$ and ${\metricForSecond[\indexGradient]}$ in a static channel, which is needed for obtaining the correct \ac{MV} probabilistically, as discussed in Section~\ref{ssec:why}.

\def\figureacc{2.9in}
\begin{figure*}
	\centering
	\subfloat[SNR is $0$~dB  ($\cardinalityOfLocalData=400$, $\numberOfEdgeDevices=50$).]{\includegraphics[width =\figureacc]{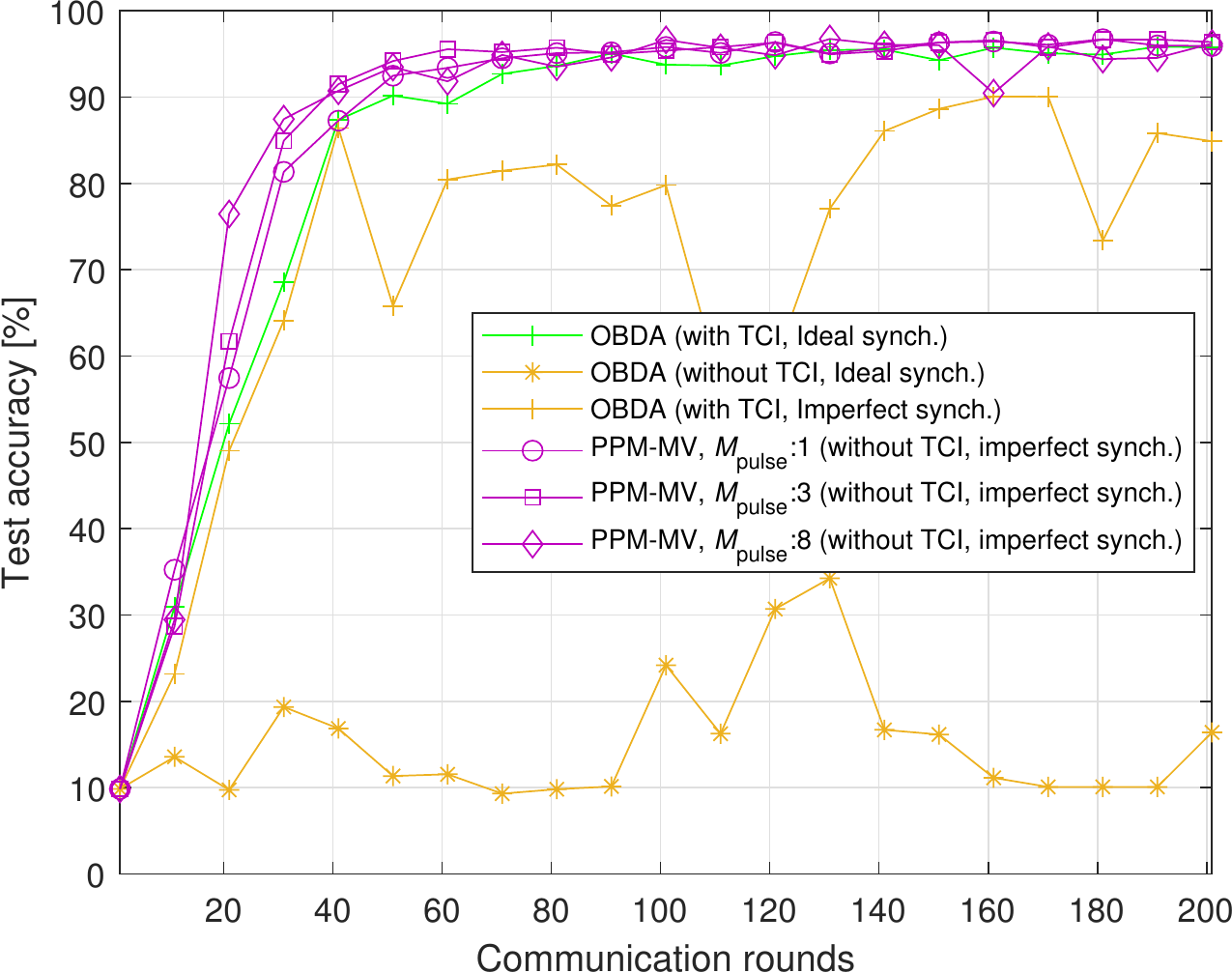}
		\label{subfig:fading0db50eds}}~~~~~~~~~~~
	\subfloat[SNR is $20$~dB  ($\cardinalityOfLocalData=400$, $\numberOfEdgeDevices=50$).]{\includegraphics[width =\figureacc]{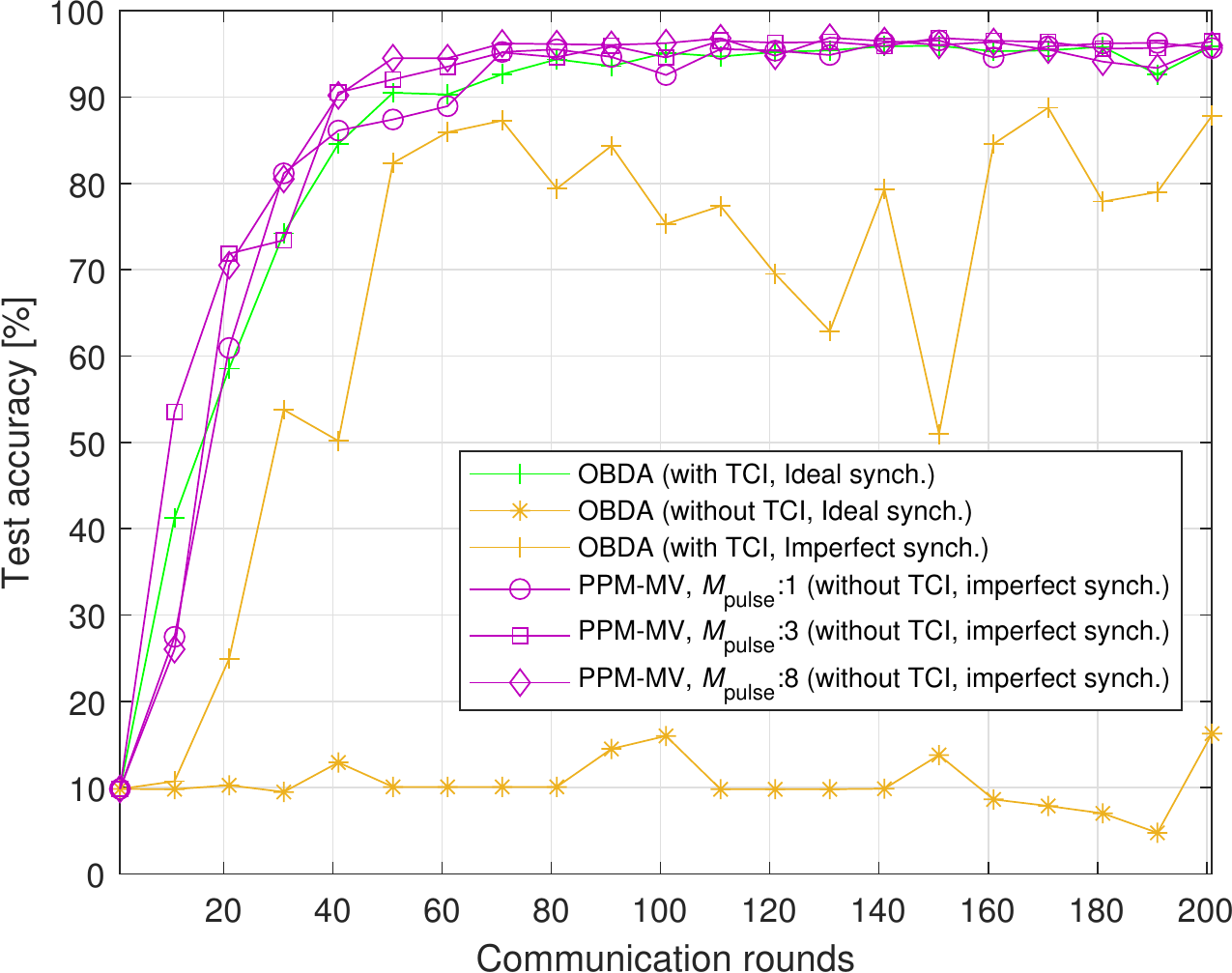}
		\label{subfig:fading20db50eds}}\\
	\subfloat[SNR is $0$~dB  ($\cardinalityOfLocalData=2000$, $\numberOfEdgeDevices=10$).]{\includegraphics[width =\figureacc]{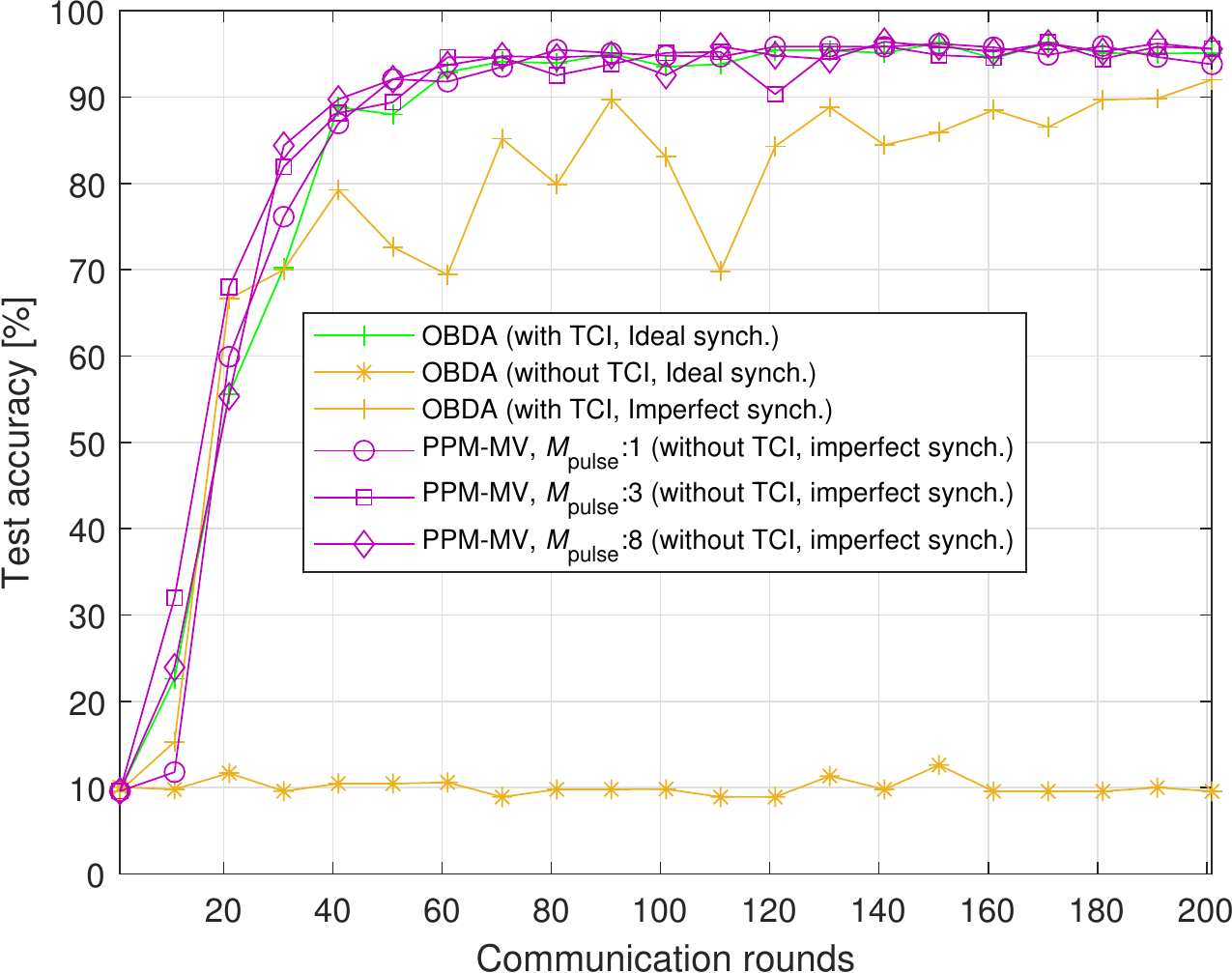}
		\label{subfig:fading0db10eds}}~~~~~~~~~~~	
	\subfloat[SNR is $20$~dB  ($\cardinalityOfLocalData=2000$, $\numberOfEdgeDevices=10$).]{\includegraphics[width =\figureacc]{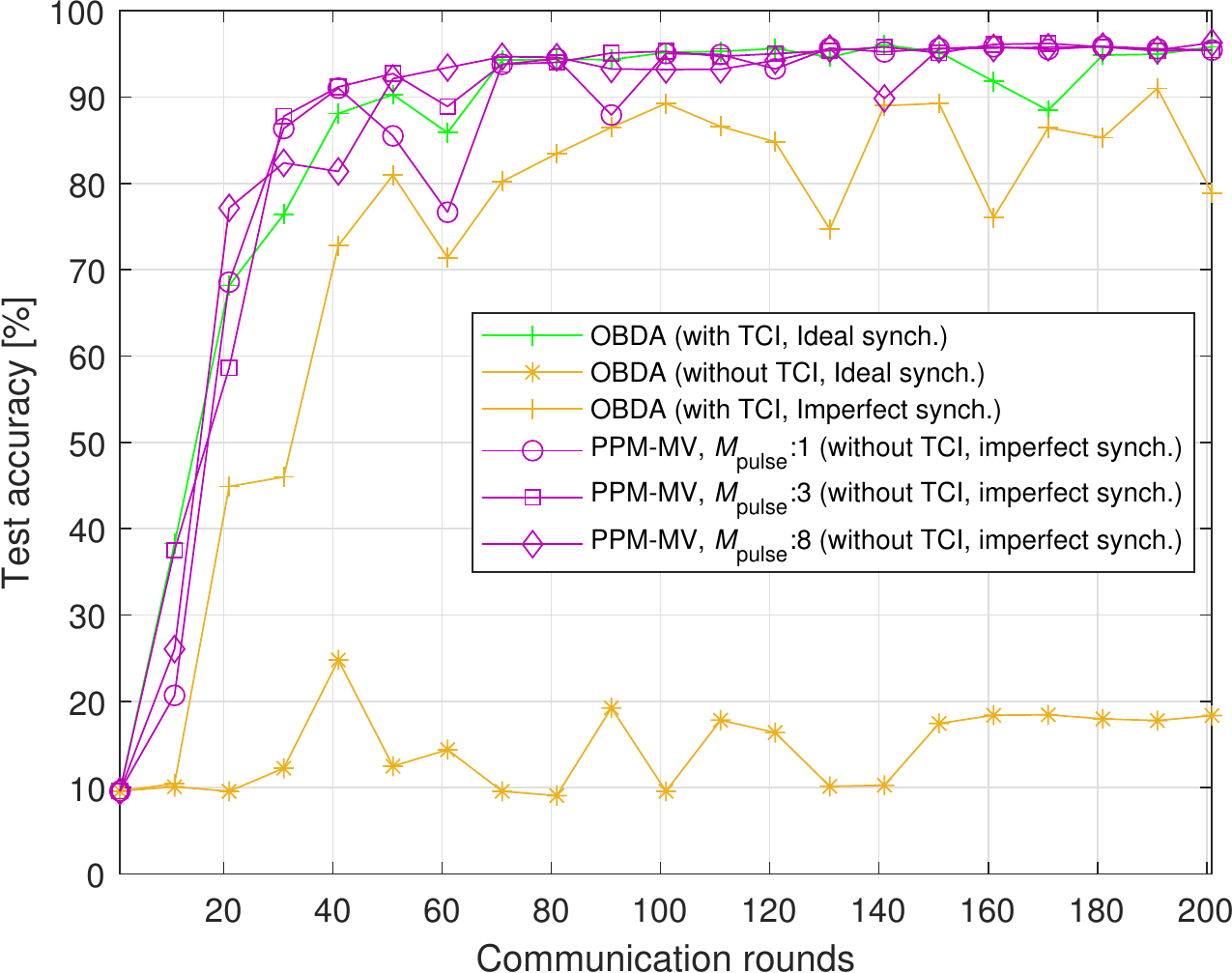}
		\label{subfig:fading20db10eds}}		
	\caption{Test accuracy results in fading channel (ITU EPA). The \ac{FEEL} with the \ac{PPM-MV} works without the  \ac{CSI} at the \acp{ED} and \ac{ES} ($\batchSize=64$).}
	\label{fig:testAcc}
	%\vspace{-4mm}
\end{figure*}

\section{Numerical Results}
\label{sec:numerical}
We consider a handwritten-digit recognition learning task over a \ac{FEEL} system, in which we compare the proposed scheme with  \ac{OBDA} \cite{Guangxu_2021}. The learning task uses the MNIST database that contains labeled handwritten-digit images of size $28\times28$, from 0 to 9.  $20000$ training images are randomly partitioned into equal shares for $\numberOfEdgeDevices\in\{10, 50\}$ \acp{ED}. We consider a \ac{CNN} for the model. It consists of one $5\times5$ and two $3\times3$ convolutional layers, each consisting of $20$ filters, and the subsequent layers to each are a batch normalization and \ac{ReLU} activation layer. Following the final \ac{ReLU} layer, a fully-connected layer of 10 units corresponding to the 0 to 9 digits and a softmax layer are utilized. %Normalization  at the input layer is not applied to the images. 
%For each update, stochastic gradient descent with a momentum of $0.9$ is applied. 
The learning rate is set to $0.01$ and $\batchSize=64$. For the test accuracy calculations, we use $10000$ test images given in the database. Our model contains $\numberOfModelParameters=123090$ learnable parameters, which corresponds to  $\numberOfOFDMSymbols=52$ \ac{OFDM} symbols for \ac{OBDA} with $\numberOfActiveSubcarriers=1200$ subcarriers. The \ac{OFDM} symbol duration $\symbolDuration$ and the threshold for \ac{TCI} are set to $66.67~\mu$s and $0.2$, respectively. To test \ac{FEEL}, two different uplink \acp{SNR} (i.e., $1/\noiseVariance$) of $0$~dB and $20$~dB are considered. ITU Extended Pedestrian A (EPA) with no mobility is considered for the fading channel, and the channels between the \acp{ED} and \ac{ES} are regenerated to capture the long-term channel variations. The \ac{RMS} delay spread of the EPA channel is $\rmsDelaySpread=43.1$~ns. As a rule of thumb, we assume that the maximum-excess delay is $\channelSpread\triangleq4\rmsDelaySpread=172.5$~ns.  We set the sample rate to  $30.72$~Msps and $\idftSize=2048$. Unless otherwise stated, we assume that the maximum synchronization error among the \acp{ED} is  $\syncError=55.6$~ns (i.e., the reciprocal of the signal bandwidth and approximately 2 samples within the \ac{CP} window). %also assume that the \ac{ES} intentionally starts the processing by backing of 4 samples in the time domain, which corresponds to $130.2$~ns, for
 We set $\uniformGap$ to $7$ to ensure that $\uniformGap\ge\ceil{(\channelSpread+\syncError)/\symbolSpacing}$ for $\symbolSpacing=55.6$~ns. The number of \ac{DFT-s-OFDM} symbols for $\numberOfElementsInAPulse=1$, $\numberOfElementsInAPulse=3$, $\numberOfElementsInAPulse=8$, and $\numberOfElementsInAPulse=13$ can be then calculated as $1642$, $2052$, $3078$, and $4108$, respectively. The simulations are performed in MATLAB.

The test accuracy  results  are provided in \figurename~\ref{fig:testAcc}. 
%For an \ac{AWGN} channel, with \acp{SNR} of $0$~dB and $20$~dB for $\numberOfEdgeDevices=10$ and $\numberOfEdgeDevices=50$, each  scheme converges and presents high test accuracy, as demonstrated in \figurename~\ref{fig:testAcc}.  In \figurename~\ref{fig:testAcc}, the fading channel is considered. 
The \ac{OBDA} provides high test accuracy in the case of ideal synchronization when \ac{TCI} presents.  This is expected because the \ac{MV} calculation requires a coherent superposition of the \ac{QPSK} symbols.  However, it completely fails in the absence of \ac{TCI}. It is also very sensitive to time synchronization errors. On the other hand, \ac{PPM-MV} results in high test accuracy without \ac{TCI} or \ac{CSI} at the \ac{ES} as it exploits non-coherent techniques for the \ac{MV} computation and immune against time synchronization errors.
%Here, we begin to see larger discrepancies in the reported accuracy across all methods. In \figurename~\ref{fig:testAccNonIid}\subref{subfig:awgn0db50edsNonIid}-\subref{subfig:awgn20db10edsNonIid}, both \ac{BAA} and \ac{PPM-MV} perform significantly better than \ac{OBDA}-\ac{QPSK} for all values of $\numberOfEdgeDevices$; however, for $\numberOfEdgeDevices=10$, \ac{PPM-MV} performs slightly worse than that of \ac{BAA}, demonstrating that for a low population, the accuracy slightly suffers. With the inclusion of the fading channel, similar results are shown. In \figurename~\ref{fig:testAccNonIid}\subref{subfig:fading0db50edsNonIid}-\subref{subfig:fading20db10edsNonIid}, the behavior is similar to that of \ac{iid} test accuracy, with the major difference being that \ac{OBDA}-\ac{QPSK} fails with and without \ac{TCI}. A unique result that appears in this case is for lower values of $\numberOfEdgeDevices$, the accuracy degrades at a faster rate for lower values of $\numberOfElementsInAPulse$ than at higher ones.
%the accuracy degrades at a faster rate than at higher $\numberOfElementsInAPulse$ values.
\begin{figure}[t]
	\centering
	{\includegraphics[width =\figureacc]{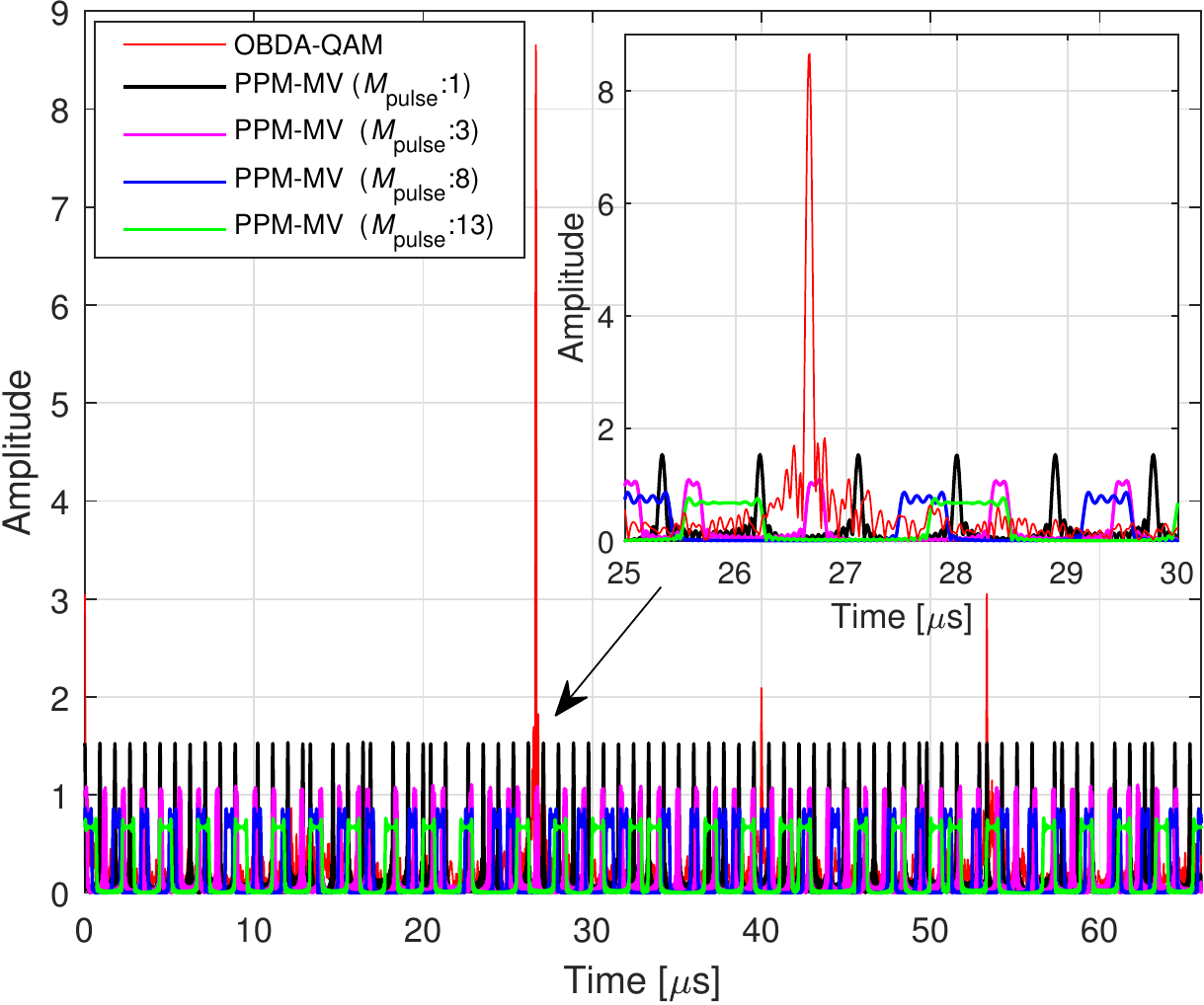}
	} 
	\caption{Temporal characteristics.}
	\label{fig:temp}
\end{figure}
\begin{figure}[t]
	\centering
	{\includegraphics[width =\figureacc]{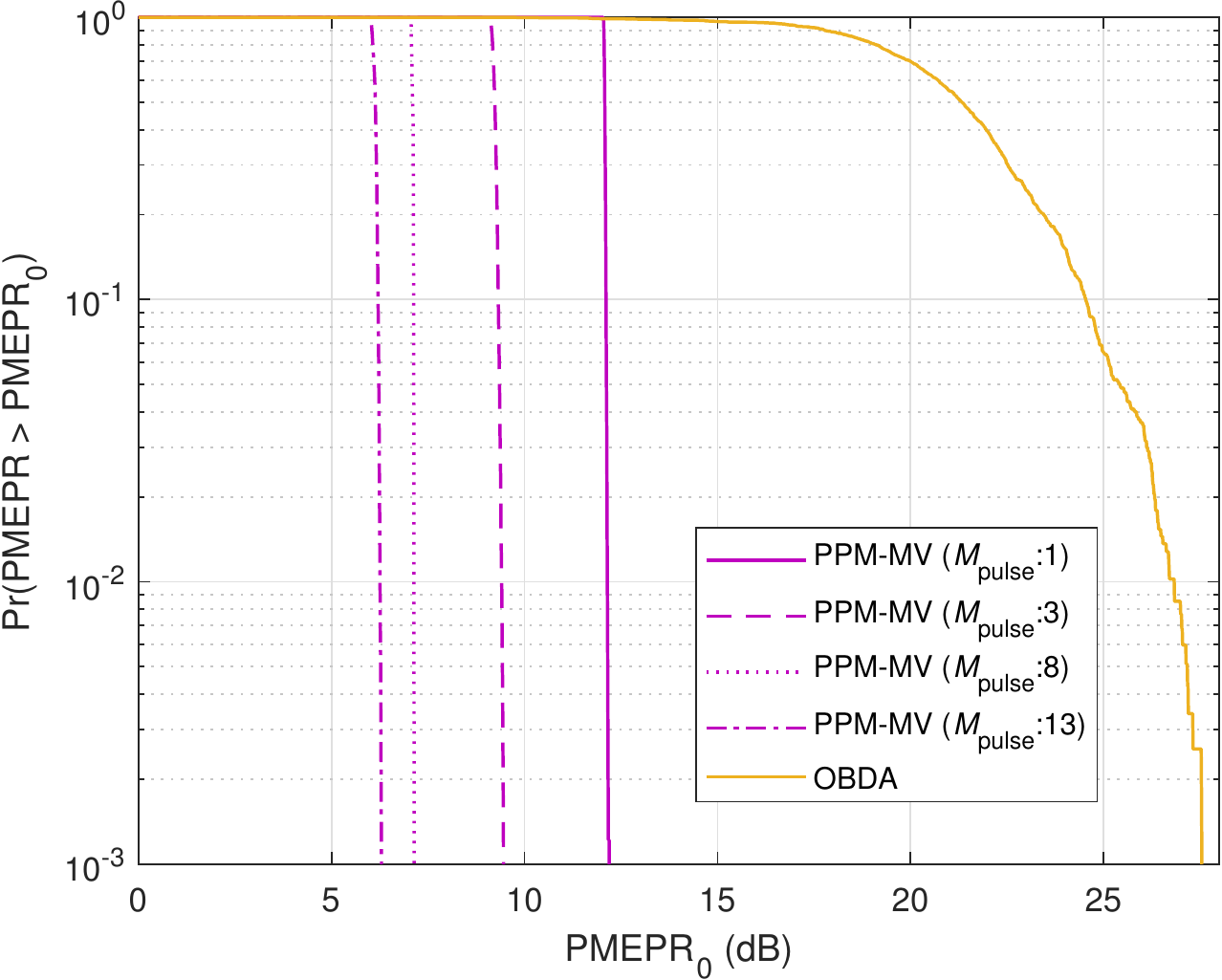}
	} 
	\caption{PMEPR distributions.}
	\label{fig:pmepr}
\end{figure}
\figurename~\ref{fig:temp} details the temporal characteristics of \ac{OBDA} and \ac{PPM-MV}. We see that the signal can be very peaky  with \ac{OBDA} when all the \ac{QPSK} symbols are similar to each other. For \ac{PPM-MV}, this is not an issue as the votes are represented as separated pulses in time. \figurename~\ref{fig:pmepr} shows the \ac{PMEPR} for both \ac{OBDA} and \ac{PPM-MV}. We show the trade-off that $\numberOfElementsInAPulse$ presents. For \ac{OBDA}, the \ac{PMEPR} can be exceptionally high. In contrast, \ac{PPM-MV} mitigates  the \ac{PMEPR} aggressively, which is an important for factor for radios equipped with non-linear power amplifiers. The trade-off displayed is that as $\numberOfElementsInAPulse$ rises, the \ac{PMEPR} curve diminishes, but, as demonstrated in \figurename~\ref{fig:temp}, more resources in time are consumed.

\section{Concluding Remarks}
\label{sec:conclusion}
Equalization methods used in traditional communications cannot be trivially used with the state-of-the-art \ac{OAC} schemes due to the superposition in the wireless channel. This issue causes a major challenge for practical \ac{OAC}. To address this problem, in this study, we propose an \ac{OAC} method that relies on non-coherent detection and \ac{PPM} symbols. We show how to design the \ac{PPM} symbols with \ac{DFT-s-OFDM} by taking time-synchronization errors into account. The main benefits of the proposed scheme as compared to the state-of-the-art solutions are that it eliminates \ac{CSI} at the \acp{ED} and \ac{ES} and provides a high test accuracy even when time synchronization is imperfect. Therefore, it offers a promising solution for enabling distributed learning over wireless networks. Also, it  reduces the \ac{PMEPR} as compared to \ac{OBDA}, where the improvement on \ac{PMEPR} can be adjusted at the expense of consuming more resources in the time domain.

\acresetall
\bibliographystyle{IEEEtran}
\bibliography{references}

\end{document}